\documentclass[12pt,a4paper]{article}
\usepackage[utf8]{inputenc}
\usepackage[T1]{fontenc}
\usepackage{amsmath,amssymb}
\usepackage{graphicx}
\usepackage{booktabs}
\usepackage{siunitx}
\usepackage[margin=1in]{geometry}
\usepackage{hyperref}
\usepackage{textcomp}
\usepackage{float}
\usepackage{caption}
\usepackage{array}
\hypersetup{colorlinks=true,linkcolor=blue,citecolor=blue,urlcolor=blue}

\begin{document}	
\begin{center}
	\textbf{{\Large Solar photocatalytic disinfection of well water using immobilized TiO$_2$:  A comparative field study with SODIS in Antananarivo}}
\end{center}

\begin{center}
	\textbf{Jean Odilon Andrianirina$ ^{1} $, Philippe Manjakasoa Randriantsoa$^{2}$,\\ Georgette Ramanantsizehena$^{3}$, Domohina Raharinirina$^{4}$, Malik Maaza$^{5}$} \vspace{0.5cm}\\
\end{center}

\begin{center}
	$ ^{1,2,3,4}$Laboratory of Matter and Radiation Physics (LPMR),\\ Department of Physics, University of Antananarivo, Madagascar \\
	$^{5}$Nanosciences and Nanotechnologies, University of South Africa \vspace{0.3cm}\\		
	\textit{Corresponding author: njakarandriantsoa@gmail.com$ ^{2}$}		
\end{center}

	\begin{abstract}
		\noindent
		Access to safe drinking water remains a major challenge in rural areas of developing countries. This study investigates the feasibility of a simple, low-cost solar photocatalytic reactor coated with a commercial titanium dioxide (TiO$_2$) pigment for the disinfection of well water contaminated with fecal coliforms. A TiO$_2$ film was deposited on a glass plate using a straightforward acetone slurry method and exposed to natural sunlight in Antananarivo, Madagascar. The efficiency was compared with the conventional SODIS method (solar disinfection without catalyst). Water samples from ten different wells were characterized for physicochemical and bacteriological quality. After 10 minutes of solar exposure, the photocatalytic reactor reduced presumptive fecal coliform counts below the method detection limit ($<$1~CFU/100~mL) in all ten samples tested, whereas the SODIS control reduced the initial count by approximately 51\% in a representative sample. While disinfection kinetics varied slightly with water turbidity and pH, inactivation below the detection limit was consistently achieved. Pearson correlation analysis showed no statistically significant linear correlation between the inactivation rate constant and any measured water-quality parameter (all $p > 0.4$). The results demonstrate that even a non-uniform, low-purity TiO$_2$ coating significantly accelerates bacterial inactivation under solar radiation, offering a promising and affordable household-scale treatment technology for low-resource settings. However, longer-term validation, catalyst stability, dark-control assessment, and leaching studies are still required before widespread implementation can be recommended.
		
		\vspace{1em}
	\end{abstract}
	
	\newpage
	
	\section{Introduction}
	
	In many rural regions of Madagascar, groundwater from shallow wells is the primary source of drinking water. However, these sources are frequently contaminated by fecal coliforms due to inadequate sanitation infrastructure and surface runoff \cite{ref1}. Conventional disinfection methods such as chlorination or boiling are often inaccessible, unsustainable, or culturally impractical for daily household use. Solar water disinfection (SODIS)---exposure of water in transparent bottles to sunlight---is a simple, low-cost technique that uses UV-A radiation and thermal effects to inactivate pathogens. Nevertheless, its efficiency is highly dependent on weather conditions and typically requires prolonged exposure (6--48 hours depending on cloud cover) \cite{ref2}.
	
	Heterogeneous photocatalysis using titanium dioxide (TiO$_2$) offers an attractive alternative to accelerate this process. When illuminated with UV light ($\lambda < 400$~nm), TiO$_2$ generates reactive oxygen species (ROS), mainly hydroxyl radicals ($\bullet$OH), which are capable of non-selectively oxidizing and destroying a wide range of microorganisms \cite{ref3,ref4}. TiO$_2$ is non-toxic, chemically stable, and can be immobilized on solid supports to facilitate reuse without the need for nanoparticle recovery. Several studies have shown that TiO$_2$ photocatalysis can drastically reduce the time required for bacterial inactivation compared to SODIS alone \cite{ref5,ref6}.
	
	While much research focuses on highly optimized anatase nanoparticles, there is a gap in translating this to real-world, low-resource fabrication. The present work aims to design a simple solar photocatalytic reactor using commercial-grade TiO$_2$ pigment (widely available and cheaper than high-purity anatase) coated on a glass plate via a rudimentary deposition method. The objective is to evaluate its efficiency for the disinfection of well water naturally contaminated with fecal coliforms. A direct comparison with the classical SODIS method is performed under identical field conditions. Ten different wells were sampled to assess the variability in local water quality and its influence on disinfection kinetics.
	
	The specific novelty of this study lies in the following: (i) the use of a commercial rutile TiO$_2$ pigment (Tionox R-937; TiO$_2$ $\ge$ 94\%, rutile $\ge$ 98\%) rather than high-purity laboratory-grade anatase; (ii) a minimal-skill deposition method that requires no specialized equipment; (iii) field validation using ten naturally contaminated well-water samples rather than synthetic laboratory water; and (iv) a direct side-by-side comparison with SODIS under identical solar conditions in Antananarivo.
	
	\section{Materials and Methods}
	
	\subsection{Sampling and water quality analysis}
	
	Water samples were collected from ten different wells in the Antananarivo region (samples S1--S10). All wells were selected based on a history of fecal contamination and proximity to potential pollution sources (latrines, runoff). Samples were taken in 0.5~L polyethylene bottles pre-rinsed with the source water, transported in a cool box at 4~$^\circ$C, and analyzed within 24~h.
	
	On-site measurements included pH (pH-meter) and electrical conductivity (conductivity meter). In the laboratory, turbidity was measured with a turbidimeter calibrated with 20--400~NTU standards. Chemical parameters (nitrates, nitrites, ammonium, calcium, magnesium, chlorides, sulfates, etc.) were determined using a photometer (Photometer 7100) following the manufacturer's protocols.
	
	\subsection{Bacteriological analysis}
	
	Presumptive fecal coliforms and total coliforms were quantified by the membrane filtration method using a Wagtech portable field kit. A volume of 100~mL of each sample was filtered through a sterile 0.45~$\mu$m membrane. The membrane was placed on a sterile absorbent pad saturated with selective culture medium (m-FC agar for fecal coliforms, incubated at 44~$^\circ$C; m-Endo agar for total coliforms, incubated at 37~$^\circ$C) for 18--24~h. Yellow colonies were counted and results expressed as colony forming units per 100~mL (CFU/100~mL). The detection limit for a 100~mL sample was 1~CFU/100~mL; therefore, ``0~CFU/100~mL'' indicates that no viable colonies were detected and that the concentration was below the method detection limit. Confirmatory tests (e.g., oxidase, Gram stain, or EC broth) were not performed due to field-kit limitations; the reported values are therefore presumptive coliform counts.
	
	\subsection{Preparation of TiO$_2$ thin film}
	
\begin{figure}[H]
		\centering
		 \includegraphics[width=0.8\textwidth]{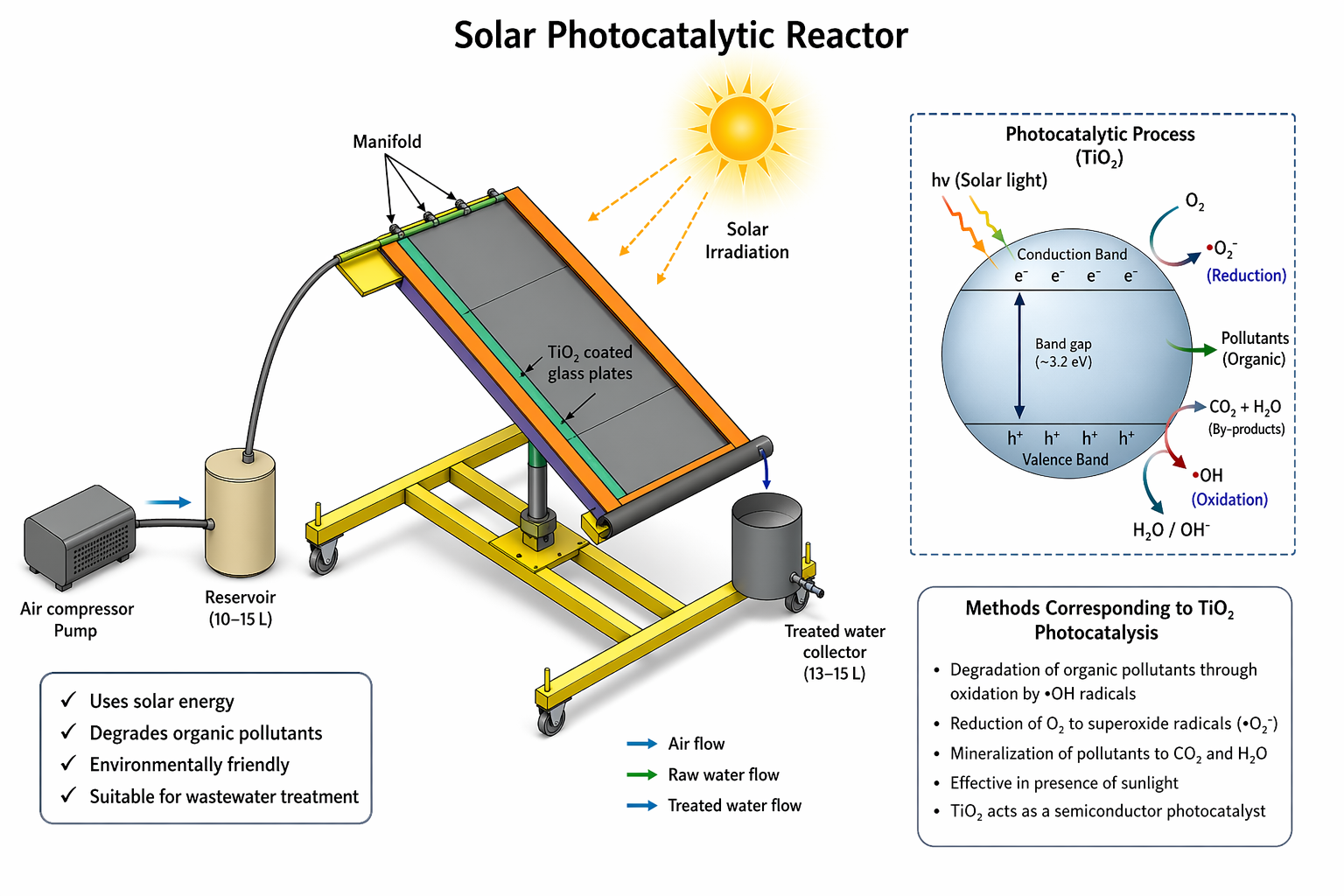}
		\caption{Schematic representation of the solar photocatalytic reactor for water disinfection using immobilized TiO$_2$. The system utilizes solar irradiation to activate the TiO$_2$ catalyst, generating reactive oxygen species ($\bullet$OH, $\bullet$O$_2^-$) responsible for bacterial inactivation.}
		\label{fig1}
	\end{figure}

	A commercial TiO$_2$ pigment (Tionox R-937; rutile structure; TiO$_2$ content $\ge$ 94\%; rutile content $\ge$ 98\%; surface treated with Al$_2$O$_3$, SiO$_2$, and organic treatment) was selected for this study due to its local availability and significantly lower cost compared to high-purity anatase nanopowders. According to the manufacturer's technical data sheet, the pigment has a pH of 6.5--7.5, oil absorption of 19--22~g/100~g, and volatile content $\le$ 0.5\% at 105~$^\circ$C. 10~g of TiO$_2$ were mixed with 250~mL of acetone to obtain a 40~g/L suspension. Glass plates (33.4~cm $\times$ 25.5~cm) were cleaned with soapy water and rinsed with hydrochloric acid. The suspension was poured onto the glass plate held at an 80$^\circ$ inclination, allowing the solvent to evaporate and the TiO$_2$ to deposit (Fig.~\ref{fig1}). The coated plates were dried in sunlight for one day, then rinsed with water to remove weakly adhered particles. The resulting coating was visibly non-uniform. The nominal TiO$_2$ loading, estimated from the suspension volume and plate area assuming 10--20\% deposition efficiency, was approximately 0.5--1.0~mg~cm$^{-2}$; because the coating was non-uniform and the loading was not gravimetrically measured, this value is indicative only. This method was chosen to evaluate the feasibility of a ``minimal-skill'' fabrication process suitable for local implementation. No XRD, SEM, particle-size, or surface-area measurements were performed in this field study; this is acknowledged as a limitation.
	
	\subsection{Solar photocatalytic experiments}
	
	Two identical reactors were used: one with a TiO$_2$-coated glass plate (photocatalytic reactor) and one with an uncoated glass plate (SODIS control). The reactors consisted of a simple metal tray (34~cm $\times$ 27~cm $\times$ 3~cm) with the glass plate placed at the bottom, providing a thin water layer of approximately 1~cm depth. For each test, 1~L of well water was poured into the reactor and exposed to direct sunlight (Fig.~\ref{fig1}). Samples were taken at 0, 5, and 10 minutes of exposure.
	
	No dark control (either TiO$_2$-coated plate in the dark or untreated water in the dark) was performed in this study. The SODIS control (uncoated glass plate under sunlight) served as the primary comparator to distinguish photocatalysis from direct solar disinfection. The absence of a dark control is acknowledged as a limitation and is recommended for future work.
	
	The experiments were performed on a clear day with ambient temperature ranging from 22--25~$^\circ$C. Although direct UV irradiance was not measured on-site due to equipment limitations, experiments were conducted between 11:00~AM and 1:00~PM local time. Based on satellite-derived climatological data for Antananarivo (approx. 18.9$^\circ$~S), the estimated ambient UV-A irradiance during the experiment was approximately 40--50~W/m$^2$, with an estimated uncertainty of $\pm$15--20\%. UV-A (315--400~nm) represents approximately 3--4\% of total solar irradiance. The primary metric for this study is the relative comparison between the TiO$_2$ reactor and the SODIS control under identical instantaneous field conditions. Water temperature during irradiation was not measured directly; this is acknowledged as a limitation.
	
	All analyses were performed in duplicate (technical replicates); results are presented as mean values. Statistical comparisons between treatments were performed using Student's $t$-test ($p < 0.05$ considered significant). For the ten samples, the apparent first-order inactivation rate constant ($k$) was estimated by fitting the data to $\ln(N_0/N) = kt$ at the 5~min time point. For samples that reached complete inactivation at 5~min ($N_5 = 0$), an exact rate constant cannot be calculated; these are reported as $k > 1.0$~min$^{-1}$. Pearson correlation coefficients between $k$ and the physicochemical parameters were calculated for the seven non-censored wells (S2, S4, S5, S6, S8, S9, S10) using standard two-tailed tests with df = 5.
	
	\section{Results}
	
	\subsection{Physicochemical characteristics of the ten wells}
	
	The physicochemical parameters of the ten samples are presented in Table~\ref{tab1}. The water was slightly acidic (pH = 5.1--6.6), with turbidity exceeding the WHO guideline ($<$5~NTU) in several wells (S7, S5, S3, S2). Conductivity was generally below 300~$\mu$S/cm, although S9 and S10 showed low values (9.8 and 9.6~$\mu$S/cm). These values were verified against the original field measurements. Ammonium was elevated in sample S4 (1.89~mg/L vs.\ guideline $<$0.5~mg/L). These values represent the initial quality before treatment.
	
	\begin{table}[H]
		\centering
		\caption{Physicochemical characteristics of the ten well water samples (S1--S10).}
		\label{tab1}
		\small
		\begin{tabular}{lcccccccccc}
			\toprule
			Parameter & S1 & S2 & S3 & S4 & S5 & S6 & S7 & S8 & S9 & S10 \\
			\midrule
			Temperature ($^\circ$C) & 20.3 & 21.3 & 21.3 & 22.5 & 21.3 & 22.0 & 22.1 & 22.0 & 22.3 & 22.5 \\
			Turbidity (NTU) & 3.8 & 6.7 & 7.2 & 1.8 & 9.3 & 0.7 & 10.3 & 3.3 & 0.7 & 0.7 \\
			pH & 5.2 & 5.3 & 6.6 & 6.2 & 5.1 & 5.2 & 5.4 & 5.8 & 5.1 & 5.1 \\
			Conductivity ($\mu$S/cm) & 49 & 208 & 139 & 289 & 99 & 129 & 112 & 218 & 9.8 & 9.6 \\
			Nitrates (mg/L) & 1.1 & 3.4 & 0.9 & 1.9 & 1.6 & 2.1 & 0.9 & 1.0 & 0.4 & 0.35 \\
			Ammonium (mg/L) & 0.01 & 0.7 & 0.03 & 2.0 & 0.02 & 0.02 & 0.12 & 0.1 & 0.02 & 0.02 \\
			Chlorides (mg/L) & 8.5 & 32 & 7.8 & 37 & 20 & 7.1 & 26.27 & 42 & 2.8 & 0.8 \\
			\bottomrule
		\end{tabular}
	\end{table}
	
	\subsection{Initial bacteriological contamination}
	
	Initial presumptive fecal coliform and total coliform counts for the ten wells are shown in Table~\ref{tab2}. All wells were contaminated, with fecal coliforms ranging from 18 to 239~CFU/100~mL and total coliforms from 6 to 239~CFU/100~mL.
	
	\begin{table}[H]
		\centering
		\caption{Initial bacteriological contamination (CFU/100~mL).}
		\label{tab2}
		\begin{tabular}{lcccccccccc}
			\toprule
			Parameter & S1 & S2 & S3 & S4 & S5 & S6 & S7 & S8 & S9 & S10 \\
			\midrule
			Fecal coliforms & 32 & 239 & 32 & 131 & 78 & 182 & 18 & 50 & 215 & 148 \\
			Total coliforms & 25 & 150 & 6 & 212 & 21 & 93 & 13 & 20 & 152 & 123 \\
			\bottomrule
		\end{tabular}
	\end{table}
	
	It is important to note that the presumptive fecal coliform counts and total coliform counts were determined using two different selective media (m-FC agar at 44~$^\circ$C and m-Endo agar at 37~$^\circ$C, respectively), each with different selectivity and recovery efficiency. The two methods are therefore not strictly nested: some thermotolerant strains that grow on m-FC agar may not grow on m-Endo agar, and vice versa. In sample S2, the presumptive fecal coliform count (239~CFU/100~mL) exceeded the total coliform count (150~CFU/100~mL). This apparent anomaly is a known limitation of field-based membrane filtration methods using different selective media, particularly when confirmatory tests (oxidase, Gram stain, or EC broth) are not performed. It does not affect the primary conclusion of this study, which is based on the relative comparison of the same medium (m-FC) before and after photocatalytic treatment. Similar discrepancies have been reported in field studies in low-resource settings \cite{ref25,ref26}.
	
	\subsection{Disinfection under TiO$_2$ photocatalysis}
	
\begin{figure}[H]
		\centering
		 \includegraphics[width=0.8\textwidth]{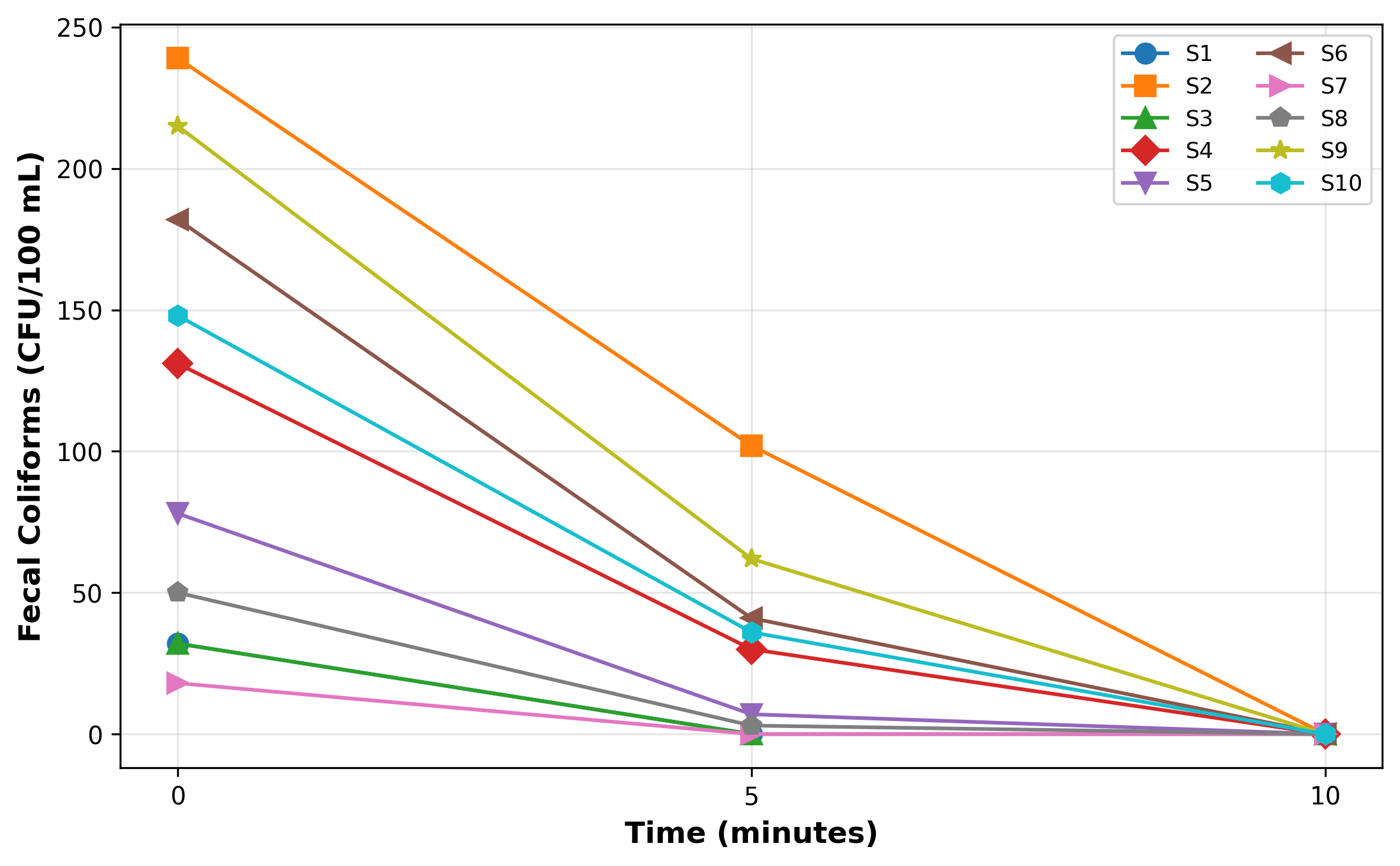}
		\caption{Disinfection kinetics of fecal coliforms for the ten wells treated with the TiO$_2$ photocatalytic reactor. All samples reached below the detection limit ($<$1~CFU/100~mL) within 10 minutes of solar exposure. Values below the detection limit are plotted at 1~CFU/100~mL with downward arrows.}
		\label{fig2}
	\end{figure}

	After 5 minutes of solar exposure, fecal coliform counts decreased substantially (Table~\ref{tab3}); some wells showed complete inactivation (S1, S3, S7). After 10 minutes, all samples reached below the detection limit ($<$1~CFU/100~mL; reported as 0~CFU/100~mL in Table~\ref{tab4}). Figure~\ref{fig2} shows the individual disinfection curves for all ten wells. For samples that reached the detection limit, values were plotted at 1~CFU/100~mL with downward arrows to indicate ``below detection limit.''
	
	Catalyst leaching during the experiments was not quantitatively evaluated in this study. This is acknowledged as a limitation and is recommended for future work.
	
	\begin{table}[H]
		\centering
		\caption{Fecal coliform counts (CFU/100~mL) after 5~min of solar exposure with TiO$_2$ photocatalysis.}
		\label{tab3}
		\begin{tabular}{lcccccccccc}
			\toprule
			Sample & S1 & S2 & S3 & S4 & S5 & S6 & S7 & S8 & S9 & S10 \\
			\midrule
			CFU/100~mL & 0 & 102 & 0 & 30 & 7 & 41 & 0 & 3 & 62 & 36 \\
			\bottomrule
		\end{tabular}
	\end{table}
	
	\begin{table}[H]
		\centering
		\caption{Fecal coliform counts (CFU/100~mL) after 10~min of solar exposure with TiO$_2$ photocatalysis.}
		\label{tab4}
		\begin{tabular}{lcccccccccc}
			\toprule
			Sample & S1 & S2 & S3 & S4 & S5 & S6 & S7 & S8 & S9 & S10 \\
			\midrule
			CFU/100~mL & 0 & 0 & 0 & 0 & 0 & 0 & 0 & 0 & 0 & 0 \\
			\bottomrule
		\end{tabular}
	\end{table}
	
	\subsection{Apparent first-order rate constants}
	
	The apparent first-order inactivation rate constants calculated at 5 minutes are presented in Table~\ref{tab5}. Wells S1, S3, and S7 achieved complete inactivation within 5 minutes, indicating very rapid kinetics ($k > 1.0$~min$^{-1}$). The remaining wells showed $k$ values ranging from 0.170 to 0.563~min$^{-1}$.
	
	\begin{table}[H]
		\centering
		\caption{Apparent first-order rate constants for fecal coliform inactivation at 5~minutes.}
		\label{tab5}
		\begin{tabular}{lcccccccccc}
			\toprule
			Well & S1 & S2 & S3 & S4 & S5 & S6 & S7 & S8 & S9 & S10 \\
			\midrule
			$k$ (min$^{-1}$) & $>1.0^*$ & 0.17 & $>1.0^*$ & 0.295 & 0.482 & 0.298 & $>1.0^*$ & 0.56 & 0.248 & 0.283 \\
			\bottomrule
		\end{tabular}
		\begin{flushleft}
			\small $^*$Complete inactivation at 5~min ($N_5 = 0$); exact rate constant cannot be calculated.
		\end{flushleft}
	\end{table}
	
	\subsection{Comparison with SODIS (sample S10)}
	
\begin{figure}[H]
		\centering
		 \includegraphics[width=0.8\textwidth]{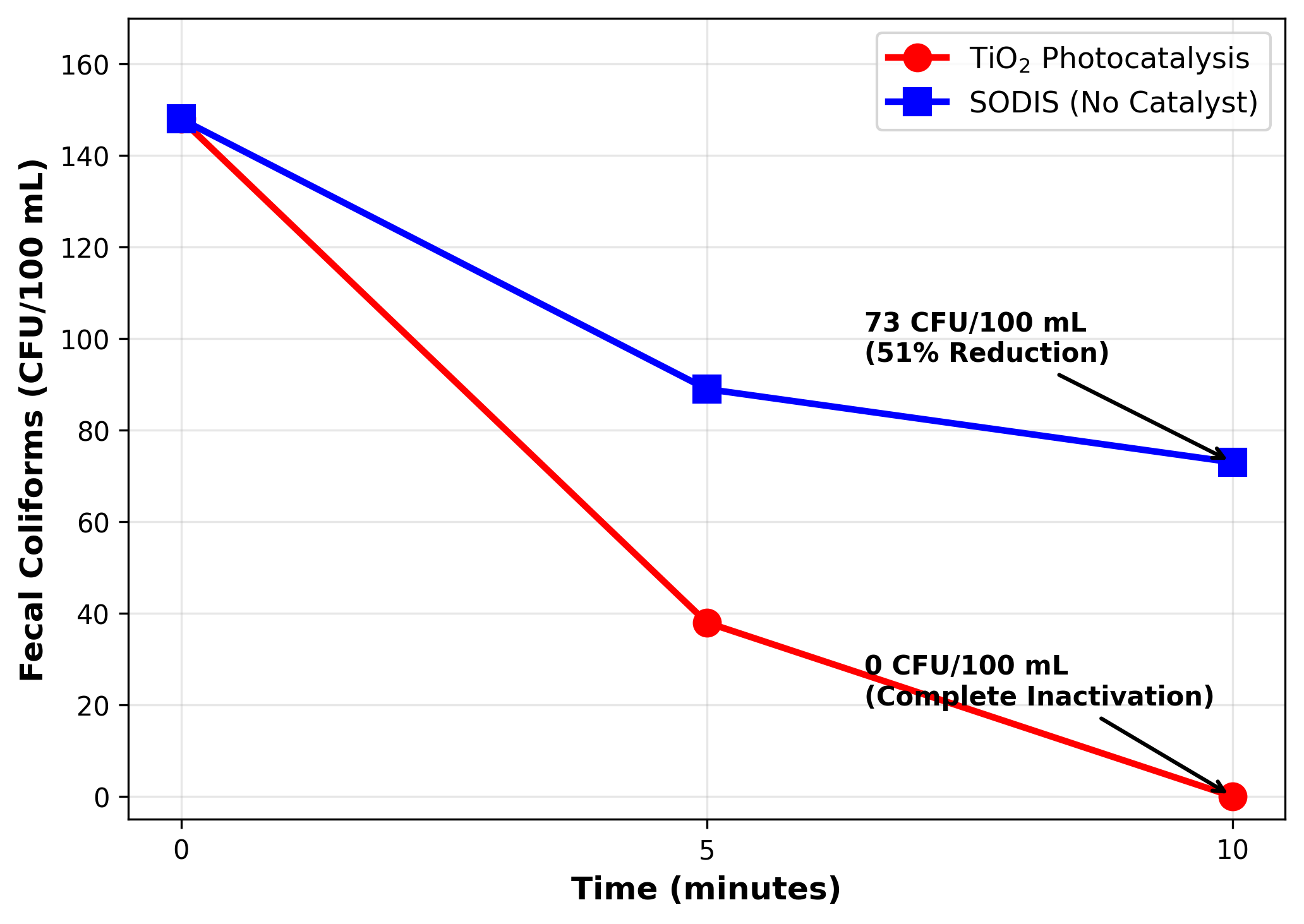}
		\caption{Comparison of fecal coliform inactivation between the photocatalytic reactor (TiO$_2$) and the SODIS control (no catalyst) for sample S10. Inactivation below the detection limit was achieved only in the presence of TiO$_2$. Values are means of duplicate measurements; duplicate agreement was within 10\%.}
		\label{fig3}
	\end{figure}

	To directly compare the photocatalytic reactor with the SODIS method, sample S10 (initial 148~CFU/100~mL) was exposed in both reactors simultaneously. Figure~\ref{fig3} shows the results. After 5 minutes, the photocatalytic reactor reduced the count to 38~CFU/100~mL (74\% reduction), while SODIS only reduced it to 89~CFU/100~mL (40\% reduction). After 10 minutes, photocatalysis achieved complete inactivation below the detection limit, whereas SODIS only reached 73~CFU/100~mL (51\% reduction). The difference at 10~min was statistically significant ($p < 0.01$, Student's $t$-test). Duplicate measurements agreed within 10\%. This confirms that the presence of the immobilized TiO$_2$ film is responsible for the accelerated kinetics. The SODIS comparison was limited to sample S10 because only one sample could be tested simultaneously under identical solar conditions during the available field window; S10 was selected as representative of mid-range contamination. A full SODIS exposure over the recommended 6--8 hours was not performed; this is acknowledged as a limitation.
	
	\subsection{Correlation with water quality parameters}
	
\begin{figure}[H]
		\centering
		 \includegraphics[width=0.8\textwidth]{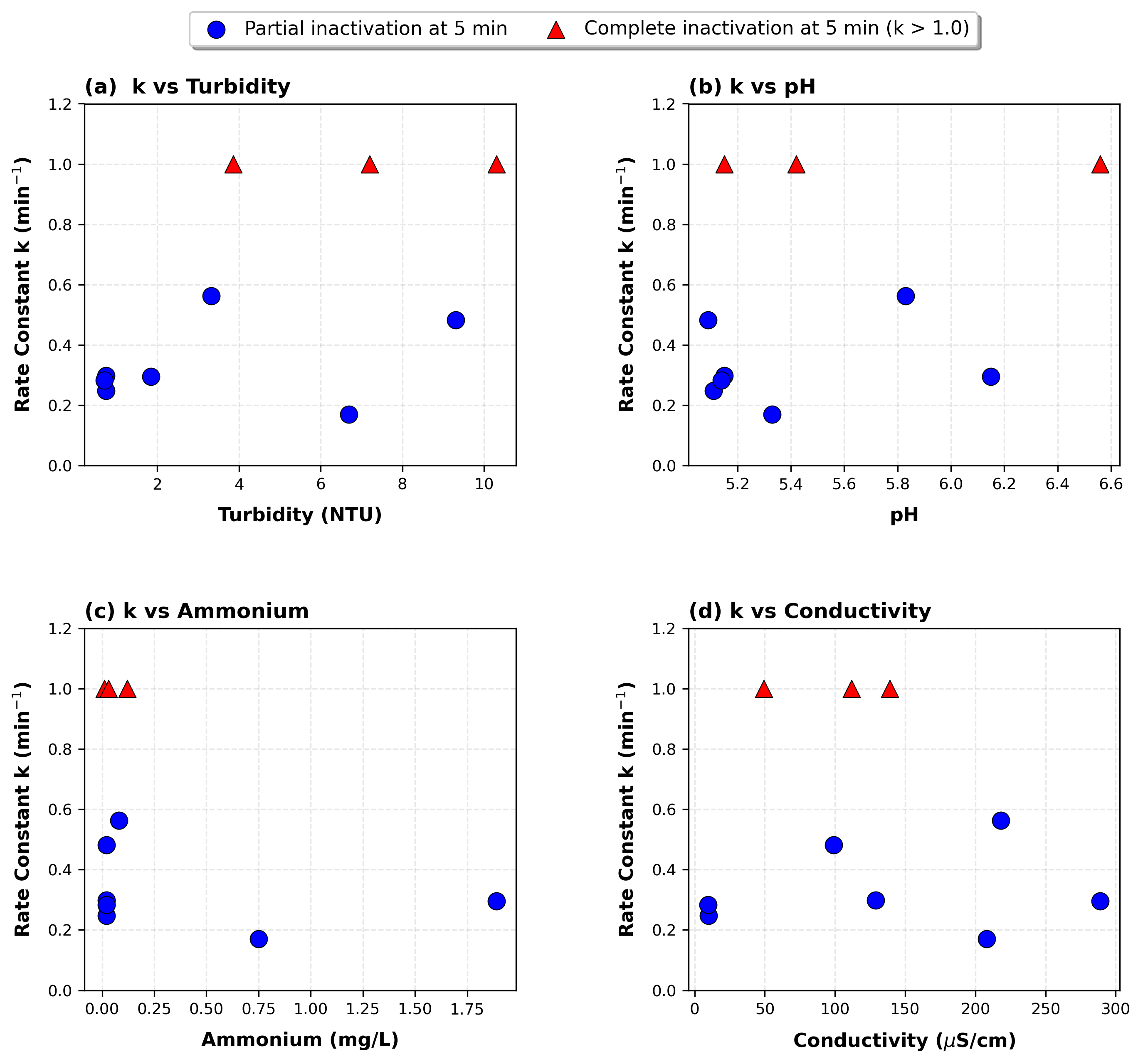}
		\caption{Correlation of the photocatalytic disinfection rate constant $k$ (calculated at 5~min) with selected water quality parameters for the ten wells. Wells S1, S3, and S7 (inactivation below detection limit at 5~min) are plotted as censored values ($k > 1.0$~min$^{-1}$) with arrows and were excluded from the quantitative Pearson correlation analysis. For the seven non-censored wells, no statistically significant correlation was observed for any parameter (all $|R| < 0.4$, all $p > 0.4$).}
		\label{fig4}
	\end{figure}

	Figure~\ref{fig4} shows the relationship between the apparent first-order rate constant $k$ (at 5~min) and key physicochemical parameters for the ten wells. Wells S1, S3, and S7 (inactivation below the detection limit at 5~min; $k > 1.0$~min$^{-1}$) are plotted as censored values and were excluded from the quantitative correlation analysis. Pearson correlation coefficients were therefore calculated for the seven non-censored wells (S2, S4, S5, S6, S8, S9, S10) and are presented in Table~\ref{tab6}.
	
	\begin{table}[H]
		\centering
		\caption{Pearson correlation coefficients between the apparent first-order rate constant $k$ (at 5~min) and selected water quality parameters for the seven non-censored wells ($n = 7$; df = 5).}
		\label{tab6}
		\begin{tabular}{lcccc}
			\toprule
			Parameter & Pearson $R$ & $R^2$ & $t$-statistic & $p$-value \\
			\midrule
			Turbidity (NTU)    & $+0.299$ & 0.089 & $+0.701$ & 0.515 \\
			pH                 & $+0.224$ & 0.050 & $+0.514$ & 0.630 \\
			Conductivity ($\mu$S/cm) & $+0.146$ & 0.021 & $+0.330$ & 0.754 \\
			Ammonium (mg/L)    & $-0.278$ & 0.077 & $-0.647$ & 0.546 \\
			Nitrates (mg/L)    & $-0.333$ & 0.111 & $-0.790$ & 0.466 \\
			Chlorides (mg/L)   & $+0.375$ & 0.141 & $+0.905$ & 0.407 \\
			\bottomrule
		\end{tabular}
	\end{table}
	
	No statistically significant linear correlation was observed for any parameter (all $|R| < 0.4$, all $R^2 < 0.15$, all $p > 0.4$). This indicates that the variability in initial counts and the presence of interfering substances did not systematically preclude complete disinfection within the 10-minute window. However, the limited sample size ($n = 7$) reduces statistical power; a weak-to-moderate correlation could remain undetected. We note that samples with higher turbidity (S7: 10.3~NTU; S5: 9.3~NTU; S3: 7.2~NTU; S2: 6.7~NTU) showed slightly lower inactivation rates at 5~min, consistent with UV light scattering and ROS scavenging effects \cite{ref17}. A larger dataset with controlled variation of individual parameters would be required to confirm these trends.
	
	\section{Discussion}
	
	\subsection{Mechanism of photocatalytic disinfection}
	
	The accelerated disinfection observed with TiO$_2$ is explained by the generation of reactive oxygen species (ROS) upon UV illumination. When a photon with energy equal to or greater than the band gap of TiO$_2$ is absorbed, an electron (e$^-$) is promoted from the valence band (VB) to the conduction band (CB), leaving a hole (h$^+$) in the valence band:
	
	\begin{equation}
		\mathrm{TiO_2} + h\nu \rightarrow e^-_{\mathrm{CB}} + h^+_{\mathrm{VB}}
	\end{equation}
	
	The photogenerated holes react with adsorbed water molecules or surface hydroxyl groups to produce hydroxyl radicals ($\bullet$OH):
	
	\begin{equation}
		h^+ + \mathrm{H_2O_{ads}} \rightarrow \bullet\mathrm{OH} + \mathrm{H^+}
	\end{equation}
	
	\begin{equation}
		h^+ + \mathrm{OH^-_{ads}} \rightarrow \bullet\mathrm{OH}
	\end{equation}
	
	Concurrently, electrons can be captured by dissolved oxygen to form superoxide radicals ($\bullet\mathrm{O_2^-}$), which can further generate hydrogen peroxide and additional $\bullet$OH:
	
	\begin{equation}
		e^- + \mathrm{O_2} \rightarrow \bullet\mathrm{O_2^-}
	\end{equation}
	
	\begin{equation}
		\bullet\mathrm{O_2^-} + \mathrm{H^+} \rightarrow \mathrm{HO_2^\bullet}
	\end{equation}
	
	\begin{equation}
		2\,\mathrm{HO_2^\bullet} \rightarrow \mathrm{H_2O_2} + \mathrm{O_2}
	\end{equation}
	
	\begin{equation}
		\mathrm{H_2O_2} + e^- \rightarrow \bullet\mathrm{OH} + \mathrm{OH^-}
	\end{equation}
	
	These ROS attack the bacterial cell membrane, leading to lipid peroxidation, disruption of membrane integrity, and eventual cell lysis. Intracellular components such as DNA and enzymes are also oxidized, resulting in irreversible inactivation \cite{ref11,ref12,ref13}.
	
	\subsection{Comparison with SODIS and the role of low-purity TiO$_2$}
	
	In the absence of a catalyst (SODIS), disinfection relies primarily on direct UV-B radiation (280--315~nm), which damages bacterial DNA, and to a lesser extent on thermal effects. The required exposure time is typically several hours \cite{ref2}. Our results show that under the same solar conditions, the TiO$_2$ photocatalyst reduced the disinfection time to 10 minutes for complete inactivation below the detection limit, whereas SODIS only achieved 51\% reduction after 10 minutes. This is consistent with previous studies that reported complete inactivation of fecal coliforms within 15--30 minutes using immobilized TiO$_2$ under sunlight \cite{ref6,ref14,ref15}.
	
	Rutile TiO$_2$, despite its narrower band gap ($\sim$3.0~eV) and generally lower intrinsic photocatalytic activity compared to anatase, has been shown to generate ROS and inactivate bacteria under UV-A irradiation \cite{ref21,ref22}. The photocatalytic activity of rutile depends strongly on its surface properties, particle size, defect structure, and the presence of surface hydroxyl groups \cite{ref23}. Several studies have demonstrated that rutile-based films can achieve significant bacterial inactivation under solar irradiation, particularly when used in thin immobilized layers where light penetration is not limited by scattering \cite{ref24,ref6}. Ohno et al.\ showed that pure rutile particles exhibit substantial photocatalytic activity for organic degradation, challenging the conventional view that rutile is photocatalytically inactive \cite{ref21}. Hurum et al.\ further demonstrated that the interfacial charge-transfer properties of rutile can contribute to overall photocatalytic efficiency in mixed-phase systems \cite{ref22}. More specifically for disinfection, Sunada et al.\ reported effective photokilling of bacteria on TiO$_2$ thin films under UV illumination, with rutile-based coatings achieving substantial inactivation within minutes \cite{ref24}. Our results are consistent with these findings and further demonstrate that a commercial-grade rutile pigment can be effective for solar disinfection in shallow water layers, likely because the immobilized configuration minimizes electron--hole recombination losses and maximizes contact between ROS and bacteria.
	
	Importantly, this study validates the use of a commercial rutile pigment rather than high-purity anatase. While anatase is generally considered more photoactive for certain applications due to its band-edge positions and surface properties, the photocatalytic activity of TiO$_2$ depends strongly on particle morphology, defects, surface area, crystallinity, and interfacial processes. For the purpose of accelerating solar disinfection in shallow water layers, our results suggest that the specific crystalline phase is less critical than the mere presence of an immobilized semiconductor surface to generate ROS. This finding has important implications for technology transfer to low-resource settings.
	
	\subsection{Influence of water matrix and practical implications}
	
	The physicochemical analysis revealed that the well water had several parameters outside the WHO drinking water guidelines, including slightly acidic pH (5.1--6.6) and elevated turbidity in some wells. Pearson correlation analysis (Table~\ref{tab6}) showed no statistically significant linear correlation between the inactivation rate constant $k$ and any individual parameter ($|R| < 0.4$; $p > 0.4$ for all six parameters), indicating that the system is robust enough to handle the typical variability of shallow well water in this region. While high turbidity (S7, S5, S3, S2) slightly slowed the initial rate---likely due to UV light scattering and competition for ROS by natural organic matter---inactivation below the detection limit was still achieved within 10 minutes.
	
	This is a critical finding for point-of-use (POU) applications: households do not need to pre-filter water to exacting standards for the reactor to work effectively within a short time window. The presence of ammonium (elevated in S4 at 1.89~mg/L) and chlorides did not appear to significantly inhibit the process, although previous studies have noted that chloride ions can act as hydroxyl radical scavengers at higher concentrations \cite{ref18}. Dissolved organic carbon (DOC/TOC) was not measured in this study; natural organic matter could influence ROS availability and photocatalytic activity, and this should be addressed in future work.
	
	\subsection{Limitations and future perspectives}
	
	This study prioritized field validation over precise material characterization. Limitations include: (i) the non-uniformity of the TiO$_2$ coating (lack of precise thickness control and surface area measurement); (ii) estimation rather than direct measurement of UV irradiance, preventing calculation of exact UV doses; (iii) single-batch testing per well without seasonal replication; (iv) duplicate technical replicates only, limiting statistical power; (v) SODIS comparison performed for only one sample; (vi) water temperature during irradiation not measured; (vii) DOC/TOC not measured; (viii) catalyst leaching and TiO$_2$ release not quantified; (ix) no dark control was performed; (x) limited sample size for correlation analysis ($n = 7$); and (xi) long-term stability and reusability not assessed.
	
	Future work should focus on: (a) performing dark controls (both TiO$_2$-coated plate in the dark and untreated water in the dark) to confirm that adsorption and natural die-off are negligible; (b) optimizing the deposition method (e.g., using a simple doctor blade technique or spray coating) to improve film uniformity and adhesion while maintaining low-cost fabrication; (c) conducting systematic studies of the influence of individual water quality parameters (pH, turbidity, NOM) under controlled laboratory conditions with larger sample sizes; (d) using a solar radiometer to quantify UV dose and enable direct comparison with other studies; (e) performing long-term stability and reusability tests over multiple cycles to assess the practical lifespan of the coated plates; (f) quantifying TiO$_2$ release into treated water and assessing potential health and environmental implications; and (g) exploring the use of visible-light-active photocatalysts or TiO$_2$ doping to extend the useful spectrum beyond UV-A \cite{ref19,ref20}.
	
	\section{Conclusion}
	
	This work demonstrates that a simple solar photocatalytic reactor using a low-cost, commercial-grade TiO$_2$ film can effectively disinfect well water contaminated with fecal coliforms in a short exposure time (10 minutes). Using ten different well water samples from Antananarivo, we showed that inactivation below the detection limit ($<$1~CFU/100~mL) is achieved consistently, despite significant variations in initial contamination levels (18--239~CFU/100~mL) and water quality parameters (pH 5.1--6.6, turbidity 0.7--10.3~NTU). Pearson correlation analysis showed no statistically significant linear correlation between the inactivation rate constant and any measured water-quality parameter (all $p > 0.4$ for the seven non-censored wells).
	
	Compared to the conventional SODIS method, which achieved only 51\% reduction after 10 minutes, photocatalysis significantly accelerates bacterial inactivation by a factor sufficient to make it a promising practical daily household intervention. The use of commercial-grade rutile TiO$_2$ pigment (Tionox R-937), rather than high-purity anatase nanoparticles, demonstrates that effective photocatalytic disinfection can be achieved with locally available, low-cost materials.
	
	Such a system represents a promising, low-cost, and sustainable household water treatment technology for rural Madagascar and similar low-resource settings. However, further validation studies---including long-term stability testing, seasonal variation, multiple-cycle reuse, catalyst leaching assessment, dark controls, and full SODIS comparison---are needed before widespread implementation can be recommended. The simplicity of the reactor design and the minimal-skill fabrication process make it well-suited for local implementation and technology transfer.
	
	\section*{Acknowledgement}
	
	The authors acknowledge the support of the University of Antananarivo for providing laboratory facilities and research assistance. We also thank Professor Malik Maaza for his valuable guidance and encouragement.
	
	\section*{AI assistance}
	
	The authors declare that artificial intelligence tools were used solely for language editing and formatting assistance. All scientific content, analysis, and interpretations were performed by the authors.
	
	\section*{Ethical approval}
	
	Ethical approval is not required for this study as it does not involve human subjects, clinical trials, or animal experimentation. The water sampling was conducted in accordance with local environmental safety guidelines.
	
	\section*{Funding}
	
	This research received no external funding.
	
	
\end{document}